\documentstyle[aps,prl,floats,graphicx]{revtex}
\begin{document}

\def\topfraction{1} \def\bottomfraction{1} \def\textfraction{0}

\twocolumn[\hsize\textwidth\columnwidth\hsize\csname @twocolumnfalse\endcsname
\title{Composite defect extends cosmology --
$^3$He analogy}
\author{V.~B.~Eltsov,$^{1,2}$ T.W.B. Kibble,$^{3}$
  M.~Krusius,$^1$ V.M.H. Ruutu,$^{1,*}$  and  G.~E.~Volovik$^{1,4}$}
\address{$^1$Low Temperature Laboratory, Helsinki University of Technology,
  P.O.Box 2200, FIN-02015
  HUT, Finland.\\
  $^2$Kapitza Institute for Physical Problems, Kosygina 2, 117334
  Moscow, Russia.\\
  $^3$Blackett Laboratory, Imperial College, London SW7 2BW, UK.\\
  $^4$Landau Institute for Theoretical Physics, Kosygina 2, 117334
  Moscow, Russia.}
\date{\today}
\maketitle
\begin{abstract}
  Spin-mass vortices have been observed to form in rotating superfluid
  $^3$He-B, following the absorption of a thermal neutron and a rapid
  transition from the normal to superfluid state. The spin-mass vortex
  is a composite defect which consists of a planar soliton (wall) which
  terminates on a linear core (string). This observation fits well within
  the framework of a cosmological scenario for defect formation, known
  as the Kibble-Zurek mechanism. It suggests that in the early Universe
  analogous cosmological defects might have formed.
\end{abstract}
\pacs{PACS numbers: 67.57.Fg, 05.70.Fh}
]

Experiments with superfluid $^3$He \cite{Ruutu,Bauerle} have shown
that quantized vortex lines are formed in the aftermath of a
neutron absorption event, during the subsequent rapid transition
from the normal to the superfluid state. These observations agree
with a theory of defect formation, the Kibble-Zurek (KZ) mechanism
\cite{Kibble,Zurek}, which was developed for the phase transitions
of the early Universe. In this scenario a network of cosmic
strings is formed during a rapid non-equilibrium second order
phase transition, in the presence of thermal fluctuations. The
real experimental conditions in the neutron irradiation experiment
of $^3$He-B (and also probably in the early Universe) do not
coincide with the perfectly homogeneous transition assumed in the
KZ scenario: The temperature distribution within the ``neutron
bubble'' is nonuniform, the transition propagates as a phase front
between the high and low-temperature phases, and the phase is
fixed outside the bubble. This requires modifications to the
original KZ scenario \cite{KV,DLZ,KT} and even raises concerns
whether the KZ mechanism is responsible for the defects which are
extracted from the neutron bubble and observed in the experiment
\cite{Ruutu2,AKV}. New measurements now demonstrate that a more
unusual composite defect is also formed and directly observed in
the neutron experiment. This strengthens the importance of the KZ
mechanism and places further constraints on the interplay between
it and other competing effects.

Composite defects exist in continuous media and in quantum field
theories, if a hierarchy of energy scales with different
symmetries is present \cite{KibbleClassification}. Examples are
strings terminating on monopoles and walls bounded by strings.
Many quantum field theories predict heavy objects of this kind,
that could appear only during symmetry-breaking phase transitions
at an early stage in the expanding Universe
\cite{HindmarshKibble,Vilenkin}.  Various roles have been
envisaged for them. For example domain walls bounded by strings
have been suggested as a possible mechanism for baryogenesis
\cite{BenMenahem}. Composite defects also provide one possible
mechanism for avoiding the monopole overabundance problem
\cite{Langacker}.

In high-energy physics it is generally assumed that composite
defects can exist after two successive symmetry-breaking phase
transitions, which are far apart in energy
\cite{KibbleClassification}. An example of successive transitions
in Grand Unification theories is $SO(10)\rightarrow SU(4)\times
SU(2)_R \times SU(2)_L \rightarrow SU(3)_C\times SU(2)_L \times
U(1)_Y \rightarrow SU(3)_C \times U(1)_Q$. In condensed matter
physics composite objects are known to result even from a single
transition, provided that at least two distinct energy scales are
involved, such that the symmetry at large lengths can become
reduced \cite{VolovikMineev}. An example is the spin-mass vortex
in superfluid $^3$He-B. It  was discovered in rotating NMR
measurements, after a slow adiabatic first order $^3$He-A
$\rightarrow$ $^3$He-B transition had taken place in the rotating
liquid \cite{Kondo}. Our new observations show that the spin-mass
vortex is also formed in a rapid non-equilibrium quench through
the second order transition from the normal phase to $^3$He-B.

Superfluid $^3$He-B corresponds to a symmetry-broken state
$U(1)\times SO(3)_L \times SO(3)_S \rightarrow SO(3)_{L+S}$, where
$SO(3)_L$ and $SO(3)_S$ are groups of rotations in orbital and
spin spaces, respectively. In this state two topologically
distinct linear defects with singular cores are possible
\cite{VolovikMineev}. Their structure can be seen from the B-phase
order parameter \cite{Vollhardt}, a 3$\times$3 matrix $A_{\alpha
j} = \Delta_{\rm
  B} \, e^{i \phi} \, R_{\alpha j} (\hat{\mathbf{n}}, \theta)$.  It is a
product of the energy gap $\Delta_{\rm B}$, the phase factor $e^{i \phi}$,
and a rotation matrix $R_{\alpha j}$.  The latter is an abstract rotation
which reflects the broken relative $SO(3)$ symmetry between spin and orbital
spaces. The unit vector $\hat{\mathbf{n}}$ points in the direction of the
rotation axis while $\theta$ is the angle of rotation.

A conventional vortex is the result of broken gauge $U(1)$
symmetry, which is common to all superfluids and superconductors.
It has $2\pi \nu$ winding in the phase $\phi$ of the order
parameter around the singular core, with integer $\nu$. This
vortex belongs to the homotopy group $\pi_1(U(1)) = Z$ and its
quantum number $\nu$ obeys a conventional summation rule $1+1=2$.
The phase winding translates to a persistent quantized
supercurrent circulating around the central core and thus this
vortex is called a ``mass vortex''.

The second type of defect appears in the order parameter matrix
$R_{\alpha j}(\hat{\mathbf{n}}, \theta)$ (Fig.~1, top). On moving
once around its core, $\hat{\mathbf{n}}$ reverses its direction
twice: First by smooth rotation while the angle $\theta$ remains
at the equilibrium value $\theta_D \approx 104^\circ$, which
minimizes the spin-orbit interaction energy. Later by increasing
$\theta$ to $180^\circ$, where both directions of
$\hat{\mathbf{n}}$ are equivalent, and then decreasing back to
$\theta_D$. The second leg in the direction reversal does not
minimize the spin-orbit interaction and hence it becomes confined
in space within a planar structure, a soliton sheet, which
terminates on the linear singular core or on the wall of the
container. This structure becomes possible through the existence
of two different energy (and length) scales: The superfluid
condensation energy defines the scale of the coherence length $\xi
\sim 10\,$--$100\,$nm, which is roughly the radius of the singular
core. The much weaker spin-orbit interaction defines the scale of
the dipolar healing length $\xi_D \sim 10\,\mu$m, at which the
angle $\theta$ becomes fixed. This length determines the thickness
of the soliton sheet. Since the matrix $R_{\alpha j}$ spans the
space $SO(3)$, this defect belongs to the homotopy group
$\pi_1(SO(3))=Z_2$, a two-element group with a summation rule
$1+1=0$ for its topological charge. Such a defect with the charge
1 is identical to its antidefect and represents a nonzero (but not
quantized) circulation of current in the spin part of the order
parameter around a singular core. It is named a ``spin vortex''.
By itself the spin vortex is an unstable structure: The surface
tension of its soliton tail leads to its annihilation.

\begin{figure}[tb]
\centerline{\includegraphics[width=\linewidth]{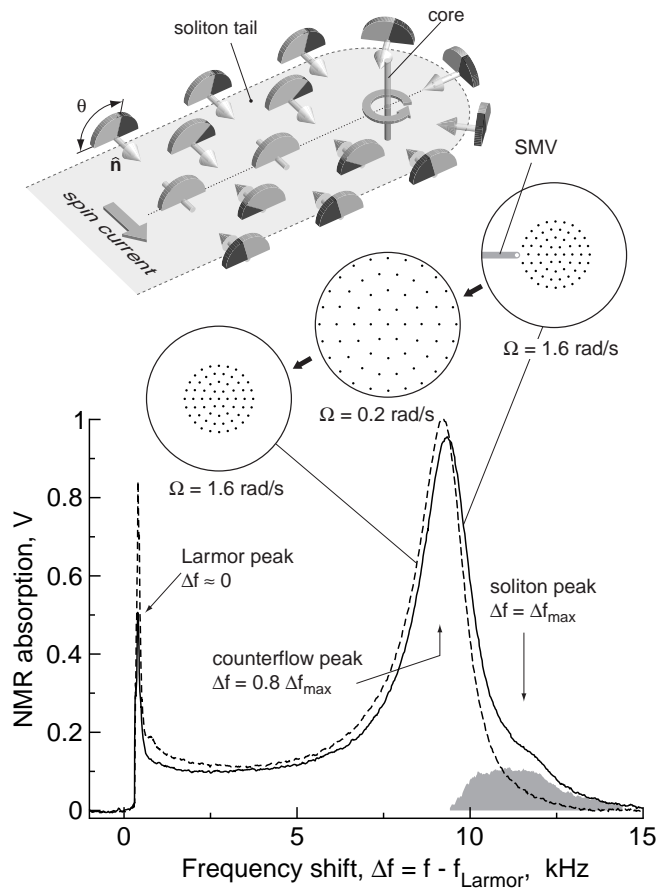}}
\medskip
\caption[]{{\em (Top)} The spin vortex in $^3$He-B is a
  disclination in the spin-orbit rotation field $R_{\alpha j}(\hat{\mathbf{n}},
  \theta)$. It
  has a singular core which is encircled by a spin current and which serves
  as a termination line to the planar $\theta$-soliton. {\em
  (Middle)} Cross sections through rotating container perpendicular to
  the rotation axis. A
  spin-mass vortex (SMV) is formed by combining a spin and mass
  vortex to a common core. Its equilibrium position is slightly
  outside the cluster of usual mass vortex lines {\em (right)}.
  By decreasing $\Omega$ to just above the annihilation threshold
  {\em (center)}, the SMV is selectively removed {\em (left)}. {\em
(Bottom)} A NMR spectrum measured
  with a SMV in the container (solid line) shows an absorption peak at the
  maximum possible frequency shift $\Delta f_{\rm max}$. This
  component in absorption originates from regions where $\hat{\mathbf{n}}$ is
  oriented perpendicular to the magnetic field
  $\mathbf H$, which occurs only in and around the soliton tail of the SMV
  (sketch on the top, $\mathbf H$ is oriented parallel to the rotation
  axis). After the SMV has been selectively removed, the only significant
change
  in the spectrum (dashed line) is the absence of the soliton contribution
  (shaded area). }
\end{figure}

Mass and spin vortices do not interact significantly -- they
``live in different worlds'', i.e. their order parameters belong
to different isotopic spaces.  The only instance where the spin
vortex has been found to remain stable in the rotating container
arises when the cores of a spin and a mass vortex happen to get
close to each other and it becomes energetically preferable for
them to form a common core. Thus by trapping the spin vortex on a
mass vortex, the combined core energy is reduced \cite{Thuneberg}
and a composite object --- $Z_2$-string + soliton + mass vortex,
or ``spin-mass vortex'' (SMV) --- is formed. Its equilibrium
position in the rotating container, which has a deficit of the
usual mass vortices, is slightly outside of the cluster of mass
vortex lines (Fig.~1, middle right). This is determined by the
balance of the Magnus force from the externally applied
normal-superfluid counterflow and the surface tension of the
soliton.

Details about the experiment are given in
Refs.~\cite{Ruutu,Ruutu2}.  The stable configuration of the
spin-mass vortex in the rotating container can be observed with
different types of NMR methods. One signature from the spin-mass
vortex in the neutron irradiation measurement is illustrated in
Fig.~2, where the height of the NMR absorption peak, used for
monitoring the number of vortex lines, is plotted as a function of
time. This accumulation record shows one oversize downward jump in
the absorption amplitude. The total number of vortex lines
accumulated by the end of the irradiation session can be
determined by different independent methods. These include: (a)
measurement of the annihilation threshold, i.e. of the rotation
velocity at which vortex lines start to annihilate at the wall of
the container during deceleration \cite{ClusterMeas}; (b)
measurement of the relative heights of the two peaks in the NMR
spectrum of Fig.~1, known as the counterflow and Larmor peaks
\cite{RatioMeas}; (c) comparison to measurements at other rotation
velocities using an empirically established $\Omega$ dependence of
the vortex formation rate \cite{Ruutu,Ruutu2}. These comparisons
prove that the large jump can include at most a few ($\lesssim 5$)
circulation quanta.

\begin{figure}[t]
\centerline{\includegraphics[width=\linewidth]{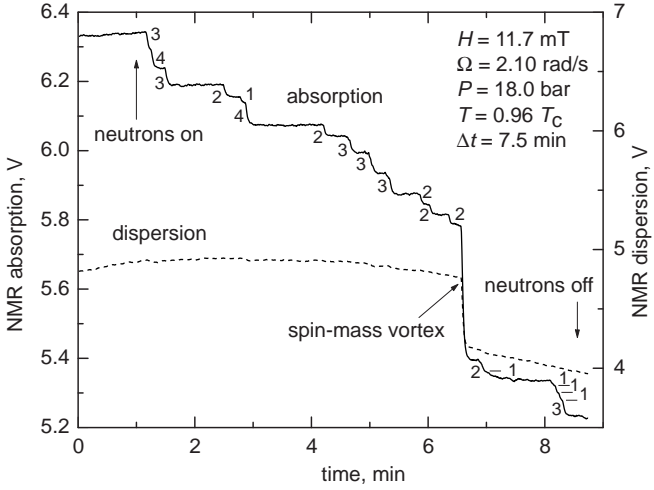}}
\medskip
\caption[]{Neutron-irradiation record of $^3$He-B and the spin-mass
  vortex. The height of the counterflow peak in the NMR absorption spectrum is
  plotted as a function of time during irradiation of the sample with
  thermal neutrons.  Each of the small downward steps in the absorption
  record marks a neutron absorption event and its height measures the
  number of newly formed mass-current vortex lines\protect\cite{Ruutu},
  also given by the number next to each step. In such events the
  (out-of-phase) dispersion signal remains unchanged (dashed line). In
  contrast the single large oversize step is recorded by both the
  absorption and dispersion signals. It is attributed to one SMV where the
  soliton tail becomes responsible for the large jumps in both signals.
  These are proportional to the length of the soliton sheet.  }
\end{figure}

A large reduction in the peak height of the NMR absorption,
together with a small number of circulation quanta, can only be
attributed to a soliton sheet which is trapped on the spin-mass
vortex. This identification is based on the change in the line
shape of the NMR spectrum with and without the spin-mass vortex
(Fig.~1, bottom). The first spectrum was recorded right after the
neutron irradiation and shows the shifted absorption at the
maximum possible frequency shift, the characteristic signature of
the soliton sheet. The second spectrum, recorded after reducing
the rotation briefly to a sufficiently low value where the
spin-mass vortex is selectively removed by pushing it to the
container wall, displays no soliton signal.  Such a recovery of
the NMR spectrum to the line shape of an axially symmetric
configuration, with a central vortex cluster surrounded by a
co-axial region of vortex-free counterflow (Fig.~1, middle left),
provides a most tangible demonstration of the initial presence of
the spin-mass vortex at the edge of the cluster. In neutron
absorption events where only mass vortex lines are formed, the
reduction in the NMR peak height is not accompanied by a frequency
shift in the location of the maximum, while the soliton produces
discontinuities in both the absorption and dispersion signals
(Fig.~2).

The spin-mass vortex is a rare product from the neutron absorption
event, compared to the yield of vortex lines. In the conditions of
Fig.~2 their ratio is roughly 1:100. Nevertheless, its presence is
thought to convey an important signal. The spin-mass vortex is the
only other type of defect, besides mass-current vortices, which so
far has been observed to form in a neutron absorption event.
(Indirect experimental evidence for the creation of $^3$He-A --
$^3$He-B interfaces has been discussed in Ref.~\cite{Ruutu2}). It
demonstrates that more than one type of order-parameter defect can
be created. This limits the possible scenarios of defect formation
which work within or around the small volume of about $100\,\mu$m
in diameter which is heated to the normal state by the energy of
the decay products from the neutron absorption reaction.  The
formation of defects occurs during the rapid cooling back to the
superfluid state on a time scale of microseconds.

At the moment the only presently viable general principle by which
defects can be created under such constraints and which would give
rise to different types of order-parameter defects is the
quench-cooling of thermal fluctuations within the KZ scenario. Two
possible routes can be suggested for the formation of the
spin-mass vortex: One possibility is that the spin and mass
vortices are formed independently, since the random
order-parameter fabric after the quench may contain discontinuity
in both the phase $\phi$ as well as in the relative rotation of
the spin and orbital axes of the order parameter.  Where these two
types of vortices happen to fuse, the combined spin-mass vortex
appears.  The long-range attractive force between the spin and
mass vortices is due to Casimir-type effects: In the vicinity of
the spin vortex the order parameter amplitude and thus the
superfluid density $\rho_s$ is reduced. This reduces the kinetic
energy of superflow $v_s$ around the mass vortex, ${1\over 2} \int
dV~\rho_s v_s^2$. The magnitude of this force is smaller by the
factor $\xi^2/d^2$ than the interaction between two vortices of
the same type, where $d$ is the distance between the~vortices.

A second possibility is a similar process as that after which the
spin-mass vortex was first observed \cite{Kondo}: In addition to
the neutron absorption event, so far the only other effective
method for forming spin-mass vortices in larger numbers is from
A-phase vortex lines when the A$\rightarrow$B transition is
allowed to propagate slowly through the rotating container. In a
neutron absorption event, AB interfaces are also among the objects
which should be formed in the KZ process
\cite{Ruutu2,Timofeevskaya}. The present measurements favor this
second explanation: In neutron absorption events spin-mass
vortices are formed only at high pressure close to the AB
transition line. Earlier measurements \cite{Ruutu2} have
established that also the yield of vortex lines from a neutron
absorption event is reduced in the vicinity of the stable A-phase
regime. The most straightforward explanation is to assume that AB
interfaces, which are formed as additional defects within the
rapidly cooling neutron bubble, intervene in the formation of
vortex lines.

Within the neutron bubble a spin-mass vortex is initially formed
as a loop which traps both the spin and mass currents, with the
soliton spanned like a membrane across the loop. This loop expands
in sufficiently strong applied counterflow to a rectilinear vortex
line, in the same manner as other loops formed from mass-current
vortices. The threshold velocity for the expansion corresponds to
the largest possible loop size, limited by the diameter of the
neutron bubble. This threshold velocity is higher for a spin-mass
vortex than for a mass-current vortex, because the energy of the
composite object is larger. In the logarithmic approximation the
energy of a spin-mass vortex is the sum of the energies of the
constituent mass and spin vortices. The energy of the spin vortex
is about 0.6 of that of the mass vortex \cite{Thuneberg}. Thus one
obtains the estimate $E_{SMV}/E_{MV}\sim 1.6$ and the same ratio
for their threshold velocities.  This is consistent with the
measurements: The threshold velocity for the creation of the mass
vortices at the experimental conditions of Fig.~2 is
$0.75\,$rad/s, while spin-mass vortices were not observed below
2\,rad/s. However, the actual threshold velocity for the SMV might
be smaller because the irradiation time at such high rotation
velocities cannot be extended indefinitely due to the rapid
accumulation of mass vortices.

In addition to the KZ mechanism also other sources might give rise
to the mass vortex lines which are counted in Fig.~2. While the
measurements clearly point to a volume effect \cite{Ruutu,Ruutu2},
recent numerical simulations of the experiment \cite{AKV} conclude
that a surface phenomenon dominates as the origin for these
directly observed vortices. Using the thermal diffusion equation
to describe the cooling neutron bubble and a one-component order
parameter in the time-dependent Ginzburg-Landau equation to model
the order-parameter relaxation, this calculation confirms the
appearance of a tangled vortex network within the bubble volume
via the KZ mechanism. However in the presence of the externally
applied counterflow from the rotation, vortex rings are also
formed on the bubble surface, due to the classical corrugation
instability of the normal/superfluid interface. This flow
instability at the warm boundary of the neutron bubble does not
require the presence of thermal fluctuations \cite{AKV} and may be
interpreted as an instability of a vortex-sheet-like intermediate
state \cite{VortSheetInst}. These vortex rings around the boundary
of the neutron bubble screen the externally applied counterflow
and allow the random vortex network inside the bubble volume to be
dissipated.

The formation of a spin-mass vortex in the flow instability seems
unlikely, since the applied counterflow does not significantly
interact with the spin degrees of the order parameter. More
likely, fluctuations must be an essential ingredient in its
formation process. Thus here again the experiment prefers the
fluctuation-dominated KZ mechanism as a more plausible
explanation. For the description of composite defects simulation
calculations should be extended to the multi-component order
parameter of $^3$He-B. However, the interplay between the KZ
mechanism within the bubble volume and the flow instability at the
bubble surface might depend on the details of the transition
process, which cannot be described by the time-dependent
Ginzburg-Landau equations, but requires a microscopic treatment of
quasiparticle dynamics in the presence of a rapidly changing order
parameter.

The observation of the spin-mass vortex, as a product from neutron
irradiation of $^3$He-B, strengthens the importance of the
fluctuation-mediated mechanisms as the source of defect formation
in non-equilibrium transitions. It shows that composite objects do
not necessarily require for their creation two phase transitions
at very different energies.

This work was carried out with support from the EU-IHP programme
and the ESF Network on Topological Defects. We thank N. Kopnin, A.
Schakel and E. Thuneberg for discussions.



\end{document}